\documentclass[12pt]{article}

\usepackage{epsf}

\begin{document}

\begin{flushright}
SLAC-PUB-8837 \\
DESY-01-061 \\
hep-ph/0105213 \\
May 2001
\end{flushright}

\renewcommand{\thefootnote}{\fnsymbol{footnote}}

\begin{center}
\vskip 4.5\baselineskip
\textbf{\Large Yet another way to measure $\gamma$}
\vskip 2.5\baselineskip
M.~Diehl\,$^{1}$ and G.~Hiller\,$^{2,}$\footnote{Work supported by
Department of Energy contract DE--AC03--76SF00515.}
\vskip \baselineskip
1. Deutsches Elektronen-Synchroton DESY, 22603 Hamburg, Germany \\
2. Stanford Linear Accelerator Center, Stanford University,
   Stanford, CA 94309, U.S.A.
\vskip 8\baselineskip
\textbf{Abstract}\\[0.5\baselineskip]
\parbox{0.9\textwidth}{We show that the CKM phase $2 \beta+\gamma$ can
be extracted from measurement of the time dependent rates in the
decays $\bar{B}^0 \to D^{(*)\pm} M^\mp$ and $B^0 \to D^{(*)\pm} M^\mp$, where 
$M=a_0$, $\pi(1300)$, $b_1$, $a_2$, $\pi_2$, $\rho_3$.  These channels
have a large asymmetry between decays of $\bar{B}^0$ and $B^0$ into
the same final state.  Even though the branching ratios are small,
their sensitivity to $\gamma$ can be competitive with decays into
$D^{(*)}$ and $(\pi,\rho,a_1)$.}
\vskip 1.5\baselineskip
\vskip 1.5\baselineskip
\end{center}

\renewcommand{\thefootnote}{\arabic{footnote}}
\setcounter{footnote}{0}


\newpage


{\textbf{Introduction.}}  
In the Standard Model of electroweak and strong interactions, $C\!P$
violation arises naturally from mixing among three generations of
quarks, encoded in the CKM matrix.  Successful to date, this highly
predictive prescription is being tested at the $B$-factories
Belle, BaBar, CLEO.
Further information will come from ongoing and future hadron collider
experiments like Run II at the Tevatron, BTeV and LHC-B.

Among the three angles of the
unitarity triangle parameterization of the CKM matrix, 
only $\beta$ has been measured so
far. The present world average is $\sin 2 \beta= 0.48 \pm 0.16$
\cite{Mele:2001pr}.  In the future this uncertainty will be greatly
reduced, to ${\cal{O}}(10 \%)$ at $B$-factories
\cite{Harrison:1998yr}, and to a few percent at BTeV and LHC-B
\cite{Hurthibus:2001}.
{}From Standard Model fits, the angle $\gamma$ is expected to be
large, $\gamma=63^{+8}_{-11}$~degrees \cite{Mele:2001pr}, in agreement
with other analyses \cite{Ali:2001hy}.

We have recently explored hadronic $B$ decays into an isospin 1 meson
with small or zero coupling to the weak vertex, either due to a small
decay constant or because it has spin greater than one
\cite{DiehlHiller:01}. We found that specific decay modes involving
the candidates
\begin{eqnarray}
\label{eq:Ms}
M=a_0(980),a_0(1450),\pi(1300),b_1,a_2,\pi_2,\rho_3 
\end{eqnarray}
are very sensitive to strong interaction dynamics, and allow one to
quantitatively test the factorization approach. In this note we point
out that the channels $\bar{B}^0 \to D^{(*)\pm} M^\mp$ and $B^0 \to
D^{(*)\pm} M^\mp$ are well suited to probe $C\!P$ violation in the
$B_d$ system. Namely, we show that the measurement of time dependent
rates in these decays can provide theoretically clean information on
$2 \beta+\gamma$.  \\


{\textbf{Time dependent rates.}}  
Aleksan et al.~\cite{Aleksan:1991ts} have proposed to obtain CKM
information from the time dependent rates of $B$-decays into non
$C\!P$ eigenstates $f$.  Each final state $f$, $\bar{f}$ can be
reached in $B$ and $\bar{B}$ decays, both directly and via meson
mixing, which leads to interference among terms with different CKM
matrix elements.  {}From measurement of all four decay rates
\begin{eqnarray}
\Gamma ( B(t) \to f ) &\sim& e^{-\Gamma |t|} \left[ \, 
\cos^2 \left(\frac{\Delta m\, t}{2}\right)
+ \rho^2 \sin^2 \left(\frac{\Delta m\, t}{2}\right)
- \rho \sin (\phi + \Delta)\, \sin (\Delta m\, t)  \,\right] ,
  \nonumber \\
\Gamma ( \bar{B}(t) \to \bar{f} ) &\sim& e^{-\Gamma |t|} \left[ \,
\cos^2 \left(\frac{\Delta m\, t}{2}\right)
+ \rho^2 \sin^2 \left(\frac{\Delta m\, t}{2}\right)
+ \rho \sin (\phi - \Delta)\, \sin (\Delta m\, t)  \,\right] ,
  \nonumber \\
\Gamma ( B(t) \to \bar{f} ) &\sim& e^{-\Gamma |t|} \left[ \,
\rho^2 \cos^2 \left(\frac{\Delta m\, t}{2}\right)
+ \sin^2 \left(\frac{\Delta m\, t}{2}\right)
- \rho \sin (\phi - \Delta)\, \sin (\Delta m\, t)  \,\right] ,
  \nonumber \\
\Gamma ( \bar{B}(t) \to f ) &\sim& e^{-\Gamma |t|} \left[ \,
\rho^2 \cos^2 \left(\frac{\Delta m\, t}{2}\right)
+ \sin^2 \left(\frac{\Delta m\, t}{2}\right)
+ \rho \sin (\phi + \Delta)\, \sin (\Delta m\, t)  \,\right] ,
\label{eq:rates}
\end{eqnarray}
one can then cleanly extract the ratio of matrix elements, the strong
and the weak phase
\begin{eqnarray}
\label{eq:quanti}
\rho \equiv \left\vert 
 \frac{{\cal{M}}(B \to \bar{f})}{{\cal{M}}(\bar{B} \to \bar{f})
                                  \rule{0pt}{1em}} 
\right\vert , \hspace{1em}
\Delta , \hspace{1em}  \phi ,
\end{eqnarray}
respectively, the latter within a discrete ambiguity.  This is true
provided that there is only one weak phase involved in the process, as
is the case for tree level dominated decays mediated by $b \to c
\bar{u} d$ and $b \to u \bar{c} d$. For simplicity, we have neglected
in (\ref{eq:rates}) effects from the width difference of the two
neutral $B_d$ mass eigenstates.  The mass difference $\Delta m$ has
been measured in the $B_d$ system as $\Delta m_d/\Gamma_d =0.730 \pm
0.029$ \cite{Groom:2000in}. We note that Eq.~(\ref{eq:rates}) is only
valid if at least one of the final state mesons has spin zero.
Otherwise, there are several helicity amplitudes, and the extraction
of the weak phase requires angular analysis \cite{Fleischer:1996ai}.

In decays $B \to \bar{f}$ with $\bar{f}=D^+ M^-$ and $M^- =
d\bar{u}$ the weak phase $\phi$ equals $-(2\beta + \gamma)$. Here, the
first term comes from $B^0$--$\bar{B}^0$ mixing and can be
cleanly measured from the $C\!P$ asymmetry in $\bar{B}^0 \to J/\Psi\,
K^0$. From the measurement of $\phi$ we can then extract $\gamma$,
modulo a discrete ambiguity.
We stress that here one does not have to rely on factorization or any
other assumption on the strong decay dynamics: 
all quantities in (\ref{eq:quanti}) can be
extracted from a fit to the rates (\ref{eq:rates}) in a model
independent way. \\


{\textbf{Large asymmetries versus large statistics.}}  
The possibility to obtain information on $\gamma$ from time dependent
studies in the decays $(\bar{B}^0,B^0) \to D^\pm (\pi,\rho,a_1)^\mp$
has been investigated in \cite{Dunietz:1998in}. Here and in the
following, $D$ stands for both $D$ and $D^*$ mesons.  Because the
branching ratios ${\cal{B}}( \bar{B}^0 \to D^+ (\pi,\rho,a_1)^- ) \sim
10^{-3}$ \cite{Groom:2000in} are large compared to those for the
$C\!P$ conjugate parent, ${\cal{B}}( B^0 \to D^{+}
(\pi,\rho,a_1)^- ) \sim 10^{-6}$, these modes are essentially
self-tagging.  Since the amplitude ratio is roughly $\rho
\simeq |V_{ub}^* V_{cd}^{\phantom{*}}|/|V_{cb}^{\phantom{*}} V_{ud}^*|
\approx 2\cdot 10^{-2}$ and the sensitivity to $\sin(\phi \pm
\Delta)$ scales with $\rho$, large data samples are required.

The situation is different for our decays $(\bar{B}^0,B^0) \to
D^\pm M^\mp$.  Unlike the
case just discussed, the hierarchy of
decay amplitudes induced by the CKM factors is removed by the small
coupling of the meson $M$ to the weak current, yielding branching
ratios for $\bar{B}^0\to \bar{f}$ and $B^0\to \bar{f}$ of the
\textit{same order of magnitude} \cite{DiehlHiller:01}.
In the case of the $a_0$ and $\pi(1300)$ this is achieved with a
decay constant of only a few MeV, so that the ratio
\begin{eqnarray}
\rho \simeq \frac{|V_{ub}^* V_{cd}^{\phantom{*}}|}{
                  |V_{cb}^{\phantom{*}} V_{ud}^*|}\, \frac{f_D}{f_M}
\end{eqnarray}
is of order one\,! This advantage is partly compensated by the fact
that the corresponding branching ratios are only $\sim
10^{-6}$ \cite{DiehlHiller:01}, so that fewer events will be available
in the analysis than for $B\to D \pi$.

To compare the sensitivity to the weak phase in the two cases we
investigate the statistical error on a suitable $C\!P$ asymmetry $A$ in
the decays (\ref{eq:rates}),
\begin{equation}
  \frac{\Delta A}{A} = \sqrt{\frac{1-A^2}{A^2 N}} \, .
\label{eq:error}
\end{equation}
As an illustration let us take a data sample of $10^8$ fully
reconstructed $B$'s. For decays into $\pi$ we then have $N = 3 \cdot
10^5$ events and an asymmetry $A \sim \rho \approx 2 \cdot 10^{-2}$,
so that $\Delta A / A \simeq 0.1$. For decays into $M$ we have instead
$N = 10^2$ and $A \sim \rho \sim 1$, giving $\Delta A / A \simeq 0.1
\sqrt{1-A^2}$. We see that the relative statistical errors are of the
same order of magnitude. This result is general, as long as $\rho\le
1$. The rate $N$ is then controlled by $\bar{B}^0 \to \bar{f}$, the
asymmetry $A$ by the amplitude ratio $\rho$, and hence the factor $A^2
N$ in (\ref{eq:error}) by $B^0\to \bar{f}$.
The latter modes, $B^0\to D^+ \pi^-$ and $B^0\to D^+ M^-$, are those
where in naive factorization the $D$ is emitted from the weak current
while $\pi$ or $M$ pick up the spectator. Here the differences between
the mesons $M$ and $\pi$ are less pronounced, and we expect branching
ratios of similar size for all mesons $M$ and $\pi,\rho,a_1$
\cite{DiehlHiller:01}.

Theoretical uncertainties in the branching ratios of $\bar{B}^0\to D^+
M^-$ decays are not small. This does not affect the extraction of
$\gamma$, but it prevents us from making accurate predictions of event
rates and asymmetries.
One source of uncertainty are the poorly known decay constants of the
charged $a_0$, $b_1$, $\pi(1300)$. We note that their
measurement in $\tau$ decays should be within reach of the $B$- and
$\tau$-charm factories. Their size is controlled by the light
quark masses, $f_{a_0, b_1} \sim m_d-m_u$ and $f_{\pi(1300)} \sim
m_d+m_u$, and various models find values in the MeV range for the
$a_0$ and $\pi(1300)$.  The branching ratios obtained in naive
factorization are then so small that factorization breaking effects
are important. We have calculated hard gluon corrections for
the corresponding decays within QCD factorization \cite{Beneke:2000ry}
and found them comparable in size to the factorizing pieces
\cite{DiehlHiller:01}. The same will hold for the $b_1$, provided that
its decay constant (on which we have not found any information in the
literature) is not larger than a few MeV. With hard gluon exchange we
can also have decays whose branching ratio is zero in naive
factorization, namely $\bar{B}^0\to D^+ M^-$ where $M$ has spin
greater than 1. Our calculation has given branching fractions $\sim
10^{-6}$ for $b_1$, $a_2$, whereas for $\pi_2$, $\rho_3$ we only found
values $\sim 10^{-9}$.  For all mesons $M$ it is however
quite possible that other contributions such as soft interactions or
annihilation are larger than the hard ones we could calculate, so that
the corresponding decays could have branching ratios above
$10^{-6}$. The $C\!P$ asymmetries would then decrease while the event
rate would go up, with a roughly constant statistical error on the
weak phase as shown above. On the other hand, non-factorizable
contributions cannot be arbitrarily large, given the success of
factorization in the decays $\bar{B}^0 \to D^+ (\pi,\rho,a_1)^-$
\cite{Beneke:2000ry}. We also recall that annihilation graphs in our
decays have the same weak phase $\phi$ as the tree level
contributions.

Since hard and soft interactions are enhanced in the decays
$\bar{B}^0\to D^+ M^-$, strong phases can be sizeable there. We have
found that the phases induced by the $\alpha_s$ corrections in QCD
factorization are indeed large, in contrast to decays into $\pi$,
$\rho$, $a_1$. As discussed in \cite{Aleksan:1991ts} nonzero phases
$\Delta$ lead to an ambiguity in extracting $\phi$. We expect however
$\Delta$ to differ among our mesons $M$, so that a combination of
channels should be able to resolve this ambiguity. Furthermore, the
strong phases $\Delta$ carry themselves important information on the
QCD dynamics in such decays.  \\


{\textbf{Time integrated measurements.}}  
To the extent that our decays are statistics limited one will probably
not be able to use the method of time integrated observables proposed
in \cite{Silva:1998cg}, where modes with hadronic decays of both $B$
mesons from the $\Upsilon(4S)$ decay are required.  We notice however
that one can extract the interference terms from (\ref{eq:rates})
while integrating over the time $t$ if each event is weighted with
$\mbox{sgn}(t)$. Instead of tracing the complete time dependence one
then only needs to know whether the corresponding decay took place
\emph{before} or \emph{after} the one that tags the flavor of the $B$
meson.  While this will in general increase the statistical error,
such an analysis may improve the systematics.  One may also use more
refined weighting factors like $\mbox{artanh}(\kappa t)$ with a
suitable constant $\kappa$. This avoids the abrupt change of the
weight at $t=0$, but does not actually lose relevant information since
the ``signal'' terms in (\ref{eq:rates}) vanish as $\sin(\Delta m\,
t)$ for $t \to 0$. \\


{\textbf{Conclusions.}}  
We conclude that, even though the decays into mesons $M$ are rare with
branching ratios $\sim 10^{-6}$, their statistical errors in the
determination of $\gamma$ are competitive with the decays into $\pi$,
$\rho$, $a_1$.  Systematic uncertainties in the two types of channels
will however be very different, so that they are indeed complementary.
We stress that there are many final states (see (\ref{eq:Ms})) where
this method can be applied. Decays involving the mesons $M$ thus open
new perspectives for the $B$ factories to perform clean and
independent tests of the CKM picture of $C\!P$ violation.  \\

{\textbf{Note added in proof:}} 
$C\!P$ violation can also be studied in charmless
$B$ decays involving the mesons $M$ discussed here. This has been
explored independently in a recent work by Laplace and Shelkov
\cite{Laplace:2001qe}.

\vspace{1cm}

{\textbf{Acknowledgments.}}  We thank R.~Fleischer, H.~Quinn and
A.~Soffer for valuable comments on the manuscript.


\end{document}